\documentclass[a4paper,envcountsame]{llncs}
\usepackage[T1]{fontenc}
\usepackage[latin9]{inputenc}
\usepackage{graphicx}

\makeatletter

\pdfpageheight\paperheight
\pdfpagewidth\paperwidth

\@ifundefined{showcaptionsetup}{}{%
 \PassOptionsToPackage{caption=false}{subfig}}
\usepackage{subfig}
\makeatother

\begin{document}

\title{LiftUpp: Support to develop learner performance}

\author{Frans A. Oliehoek\inst{3} \and
 Rahul Savani\inst{3}\and 
 Elliot Adderton\inst{2}\and 
 Xia Cui\inst{3} \and
 David Jackson\inst{3} \and 
 Phil Jimmieson\inst{3} \and 
 John Christopher Jones\inst{1} \and 
 Keith Kennedy\inst{4} \and 
 Ben Mason\inst{1} \and 
 Adam Plumbley\inst{3} \and
 Luke Dawson\inst{2} 
}
\institute{
 LiftUpp Limited \and 
 School of Dentistry, University of Liverpool \and 
 Department of Computer Science, University of Liverpool \and
 Clinical Trials Research Centre, University of Liverpool
}

\maketitle

\begin{abstract}
Various motivations exist to move away from the simple assessment of knowledge towards the more complex assessment and development of competence. However, to accommodate such a change, high demands are put on the supporting e-infrastructure in terms of intelligently collecting and analysing data. In this paper, we discuss these challenges and how they are being addressed by LiftUpp, a system that is now used in 70\% of UK dental schools, and is finding wider applications in physiotherapy, medicine and veterinary science. We describe how data is collected for workplace-based development in dentistry using a dedicated iPad app, which enables an integrated approach to linking and assessing work flows, skills and learning outcomes. Furthermore, we detail how the various forms of collected data can be fused, visualized and integrated with conventional forms of assessment. This enables curriculum integration, improved real-time student feedback, support for administration, and informed instructional planning. Together these facets contribute to better support for the development of learners' competence in situated learning setting, as well as an improved experience.  Finally, we discuss several directions for future research on intelligent teaching systems that are afforded by using the design present within LiftUpp.
\end{abstract}

\section{Introduction}

It is widely accepted that improved employability and salary prospects
are key drivers for undertaking higher education. Ensuring employability
requires that universities develop \textquoteleft highly skilled graduates
who are able to respond to the ever changing and complex needs of
the contemporary workplace\textquoteright{} \cite{andrews2008graduate}.
The importance of this link between university education and employability
is championed by the UK government, with employment as one of the
key evaluation metrics for the Teaching Excellence Framework \cite{TEF},
which also stresses the importance of \textquoteleft Stretch\textquoteright{}
(standards and assessment being effective in stretching students to
develop independence), and \textquoteleft Personalised Learning\textquoteright{}
(students\textquoteright{} academic experiences being tailored to
the individual, maximising rates of retention, attainment and progression).
Unfortunately, traditional approaches to university education are
focused on students gaining and demonstrating knowledge, and often
little emphasis is placed on its contextual application to the real-world. 

An exception is within medical programmes, especially in dentistry,
where the regulating bodies, such as the General Dental Council (GDC)
in the UK, require all graduates to be appropriately competent at
the point of graduation \cite{council2011preparing}. In this context,
competency refers to the \textquoteleft ability (of the learner) to
adapt and to flexibly apply and develop knowledge and skills in the
face of evolving circumstances\textquoteright{} \cite{govaerts2013validity}
across multiple domains (Professionalism, Management \& Leadership,
Knowledge, Communication, Clinical). However, to support the development
of competency, new and sophisticated methods of assessment that are
seamlessly integrated with curriculum design are required.  

This need for increased sophistication and integration puts high demands
on the e-infrastructure because it requires more intelligent systems
that can be used in the work place every day, and that can fuse all
forms of assessment data together. These systems must also be able
to enhance student development through personalised real time feedback
to drive changes to learner self-regulation. Moreover, the information
these systems provide must be robust and defensible to enable safe
decisions on student progress in dentistry education in order to protect
the public \cite{Dawson16calling}.

In this paper, we describe LiftUpp, a system developed at the School
of Dentistry at the University of Liverpool, which has been specifically designed to
meet these demanding requirements. Therefore, LiftUpp offers the opportunity
to spawn cutting edge approaches to Artificial Intelligence in Education
(AIED), which can be applied to all (including clinical) subjects.

\section{Background and Related Work}

Over the last decades, in dentistry, as well as other fields of higher
education, more focus has been put on graduate employability. In terms
of Miller's pyramid \cite{Miller90assessment}---which divides learner
competence into 4 levels of increasing mastery: \emph{knows}, \emph{knows
how}, \emph{shows how }and \emph{does}---this means that a strong
emphasis needs to be put on the higher levels of the hierarchy. Moreover,
it is important to assess and stimulate the improvement of student
\emph{performance} over time:
\begin{quote}
`Competence indicates what people can do in a contextual vacuum,
under perfect conditions. This might be evident using controlled assessment
methods looking at the lower tiers of Millers pyramid. Performance,
however, indicates how people behave in real life, on a day-to-day
basis.' \cite{urlWBA}
\end{quote}
One school of thought is that a critical pedagogy for developing performance
in students is \emph{deliberate practice}: learners are systematically
challenged by selecting tasks of increasing difficulty, particularly
targeting those aspects that the learner needs to improve. This requires
high quality multisource feedback (in dentistry education, for example,
from both staff and patients \cite{Dawson16calling}). In other words,
in moving towards educational systems that focus on employability,
it is needed to (1) have accurate skill or \emph{performance} assessment,
(2) identify the best tasks to stimulate learning, (which depends
on assessment as well as on models of student learning), (3) provide
the right feedback at the right times. 

We believe that this presents a huge opportunity for applying techniques
from the field of Artificial Intelligence in education (AIED). For
instance, data-driven models of student learning and knowledge assessment
\cite{Koedinger13AIMag,MVGB00,YKG13} directly target need (1) and
better ways of providing feedback \cite{VanLehn06IJAIE} or hints
\cite{Barnes10ETS} or adaptive selecting next assignments \cite{Rafferty11AIED}
fall squarely in research on \emph{intelligent tutoring systems (ITSs).
}The potential benefits of ITSs in more conventional teaching are
clearly demonstrated by a number of systems, such as the Cognitive
Tutor \cite{Ritter07Pschyco}, which led to better student learning
in a Algebra course. 

However, in dentistry education a large part of the assessment is
\emph{workplace based}, i.e., consists of actual treatments that are
(under supervision) performed on real patients. In particular, in
the School of Dentistry at the University of Liverpool, students need to train to develop
skills for a fixed, but large set of \emph{procedures }(i.e., particular
dental treatments). These procedures are trained for in clinical \emph{sessions}
(i.e., one training attempt for a particular procedure). These sessions
are observed by \emph{observers, }instructors that both assesses the
performance and that step in when needed. The number of sessions that
a student takes for a particular procedure is not fixed; they need
to demonstrate that they can perform the procedure consistently well.
Such workplace-based assessment settings are difficult to address
with existing AIED techniques due to a few main reasons.

First, most existing techniques have been operationalised in much
simpler settings, thus allowing the imposition of strong assumptions
underlying the proposed methods. For instance, the Cognitive Tutor
is based on the theory of ACT-R \cite{Anderson98ACTbook}, which assumes
that skills and knowledge can be decomposed into so-called \emph{knowledge
components (KCs)}, which can subsequently be tracked (frequently referred
to as\emph{ knowledge tracing }\cite{Corbett95UMUAI}).\emph{ }Ritter
et al.~\cite{Ritter07Pschyco} were able to come up with a suitable
decomposition of their algebra domain, but in workplace-based settings
this might be much more difficult.

A second reason, and direct consequence of the first, is that, for
most approaches, the data collection is more straightforward. For
instance, the learner models examined by Rafferty et al. \cite{Rafferty11AIED}
are trained on sequences of data of the form \{<exercise1, correct\_answer1>,
<explain\_action>, <exercise2, incorrect\_answer2>, ...\}. Here, each
observation in the sequence directly corresponds to a possible transition
of the learners knowledge state (in case of an explain action) or
a observation of it (in case of an exercise). Given that the definition
of KCs is much more difficult in workplace based settings, the question
of what data to collect (e.g., at what resolution) and how to collect
it become difficult. 

Third, educational programs with a workplace based component inherently
lead to multi-source data, which can significantly complicate fusing
the resulting data. Moreover, as we argued, the information that we
want to infer should relate to predictions of performance over time,
which is arguably more complex than just trying to predict the current
state of a number of KCs in a learner.

These complications present large challenges for the AIED community.
In the remainder of this paper we present aspects of LiftUpp that
make a first step in addressing some of these challenges. In particular,
LiftUpp collects data on performed treatments in terms of what can
be observed not what constitutes jumps in learner knowledge, thus
bypassing the question of KC definition. It also operationalises this
large-scale data collection problem in an effective way. As such,
the system establishes an approach for data collection that can serve
as a starting point for alternatives to knowledge tracing, such as
the recently proposed variant of knowledge tracing based on deep learning~\cite{Piech15NIPS},
that are less dependent on particular properties of structured domains.
LiftUpp also includes (preliminary) methods and metrics for assessing
learner performance over time in workplace-based teaching. In these
steps to address the above challenges lies the value of the system
for the AIED community.

\section{Overview of the LiftUpp Platform }

LiftUpp is a digital educational platform designed to support quality-assured
assessment, feedback, curriculum design and mapping. Its design is
grounded in pedagogy and directly addresses the issues of complex
data collection, triangulation and integration with stakeholder outcomes,
as well as the provision of detailed feedback to drive changes to
learner self-regulation through engendering the right educational
impact, and it clears the way for applying AI and data-driven improvements
to workplace-based education.\textbf{ }It is the most sophisticated
digital educational platform for workplace-based assessment used in
dentistry, and is currently deployed in 70\% (10 out of 14) of UK
dental schools, \textbf{}as well as in veterinary medicine, physiotherapy,
nursing and other healthcare courses.\textbf{}

\begin{figure}[t]
\begin{centering}
\includegraphics[width=10cm]{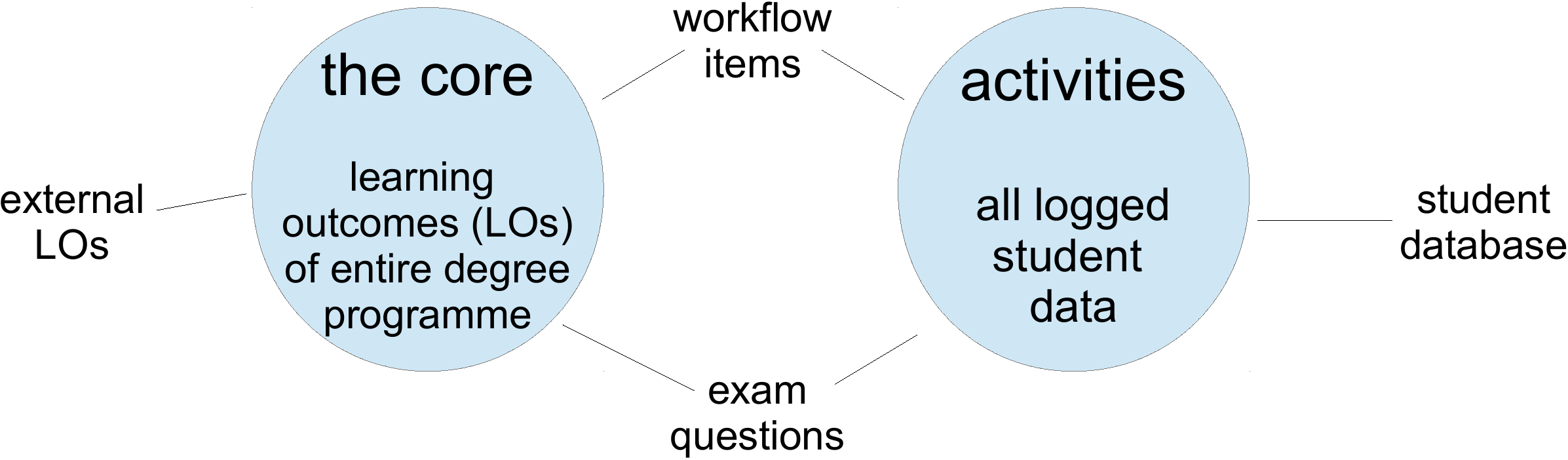}
\par\end{centering}

\protect\caption{A graphical overview of the Core of LiftUpp.}

\label{fig:ANONSYS-overview} 
\end{figure}

\begin{figure}
\begin{centering}
\includegraphics[width=1\columnwidth]{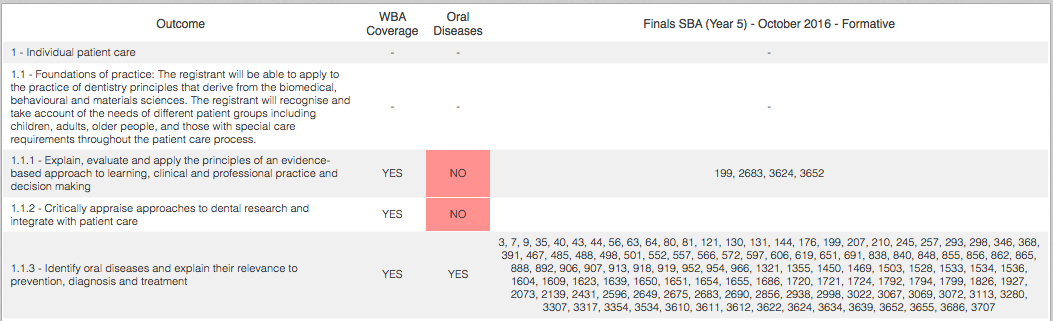}
\par\end{centering}

\protect\caption{The mapping data returned when the system is asked \textquoteleft show
me the workbased (WBA) coverage, Oral diseases teaching, and questions
used in the final examination (Oct 2016) for each GDC learning outcome\textquoteright .}

\label{fig:mapping-example} 
\end{figure}

An overview of the LiftUpp platform is given in Fig.~\ref{fig:ANONSYS-overview}.
The figure shows the \textquoteleft core\textquoteright , which contains
the learning outcomes of the entire program (both internal as well
as those of external stakeholders such as accreditation bodies), and
connects them to exams, and, in fact, individual exam questions and
\emph{work flow items} (discussed further below), of treatments that
the students need to perform.  Sitting off the core are several modules,
which currently comprise: an assessment building module (with support for exam setting, QA,
blueprinting, psychometrics, reviewing, results, feedback); an iPad-based
data collection module; and a web portal (system administration, data
analysis, collation and display). Each of these modules is mapped to the core in a one to one or one
to many relationship, which enables every item of collected data to
be instantly mapped, blueprinted, and integrated to any stakeholder,
any domain, and any metadata with which it is collected such as location,
staff member, patient etc.\textbf{ }For instance, Fig.~\ref{fig:mapping-example}
shows an example how the part of the curriculum can be investigated
in the context of learning outcomes. 

This mapping of every single element of the entire degree program
to the desired learning outcomes is what enables new forms of interpretation
of the performance and competence of learners against the set of learning
outcomes. We will discuss such novel forms of interpretations and
resulting ways of supporting the development of learners in Section~\ref{sec:data-fusion-and-use}.
First, however, we discuss the collection of data in Section~\ref{sec:data-collection}.

\section{Data Collection for Workplace-based Education}

\label{sec:data-collection}

While storing assessment results of students is nothing novel in principle,
a salient feature of LiftUpp is the level at which these are recorded,
and their ability to be integrated together against each learning
outcome. This opens the way for LiftUpp to be the first platform capable
of full programmatic assessment, which is the ultimate expression
of assessment for learning\textbf{ }\cite{Vleuten1996assessment},
where all assessments are deliberately designed and integrated to
both develop and demonstrate each learning outcome in a student-centred
manner. In the extreme, the importance of individual assessments themselves
melt away and just supply data on learner performance with respect
to the learning outcomes. In this paradigm, progression is based on
performance stability and not on passing single tests, making it much
better aligned to the needs of the real-world workplace \cite{Dawson16calling}. 

However, to realise this, one significant challenge lies in effectively
managing data components from multiple sources. For the use of LiftUpp
in dentistry, this required the collection of daily observational
data from 300 students in the workplace in 20 different sites, from
100 different staff, spanning 165 learning outcomes, along with data
from other forms of assessment. While this is challenging by itself,
it is further complicated by the inherent difficulty of objectively
assessing the quality of treatments performed by students.

\paragraph{Observations using Work Flows.}

\begin{figure}[t]
\begin{centering}
\includegraphics[width=1\columnwidth]{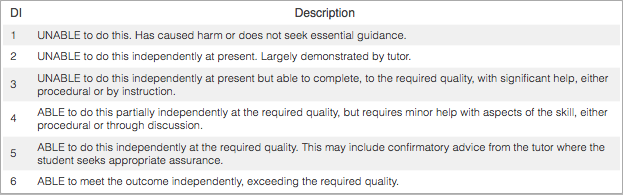}
\par\end{centering}

\protect\caption{The 6-point scale employed in LiftUpp.}

\label{fig:6pt_scale} 
\end{figure}

To enable objective assessment, LiftUpp uses the combination of a
6-point scale and a work flow model of data collection. Within the
6-point scale, shown in Fig. \ref{fig:6pt_scale}, each point is known
as a \textquoteleft developmental indicator\textquoteright . The scale
is based on evidence that suggest objectivity is enhanced when scales
are grounded in the developing the independence of the learner\textbf{
}\cite{crossley2011good}\textbf{}. 

Data supports the validity and reliability of this scale~\cite{anon}\textbf{.}
The work flow model orders the required observations into a work flow
order making it simple for the staff to observe the maximum amount
of data in real time. A key feature of the approach is that is not
an assessment, i.e., staff do not stand and watch everything and \textquoteleft tick
boxes\textquoteright . The approach is one of observation and feedback,
where staff only record what they see. Data suggest that the approach
is both efficient and acceptable to staff, and the average member
of staff is able to make 18 observations per session, per student
(staff:student 1:8). This translates to the average student having
5061 observations from 52 different members of staff by the time of
graduation.

\paragraph{Recording Performance.}

\begin{figure}
\subfloat[iPad interface for judging a clinical session.]{\begin{centering}
\includegraphics[width=6cm]{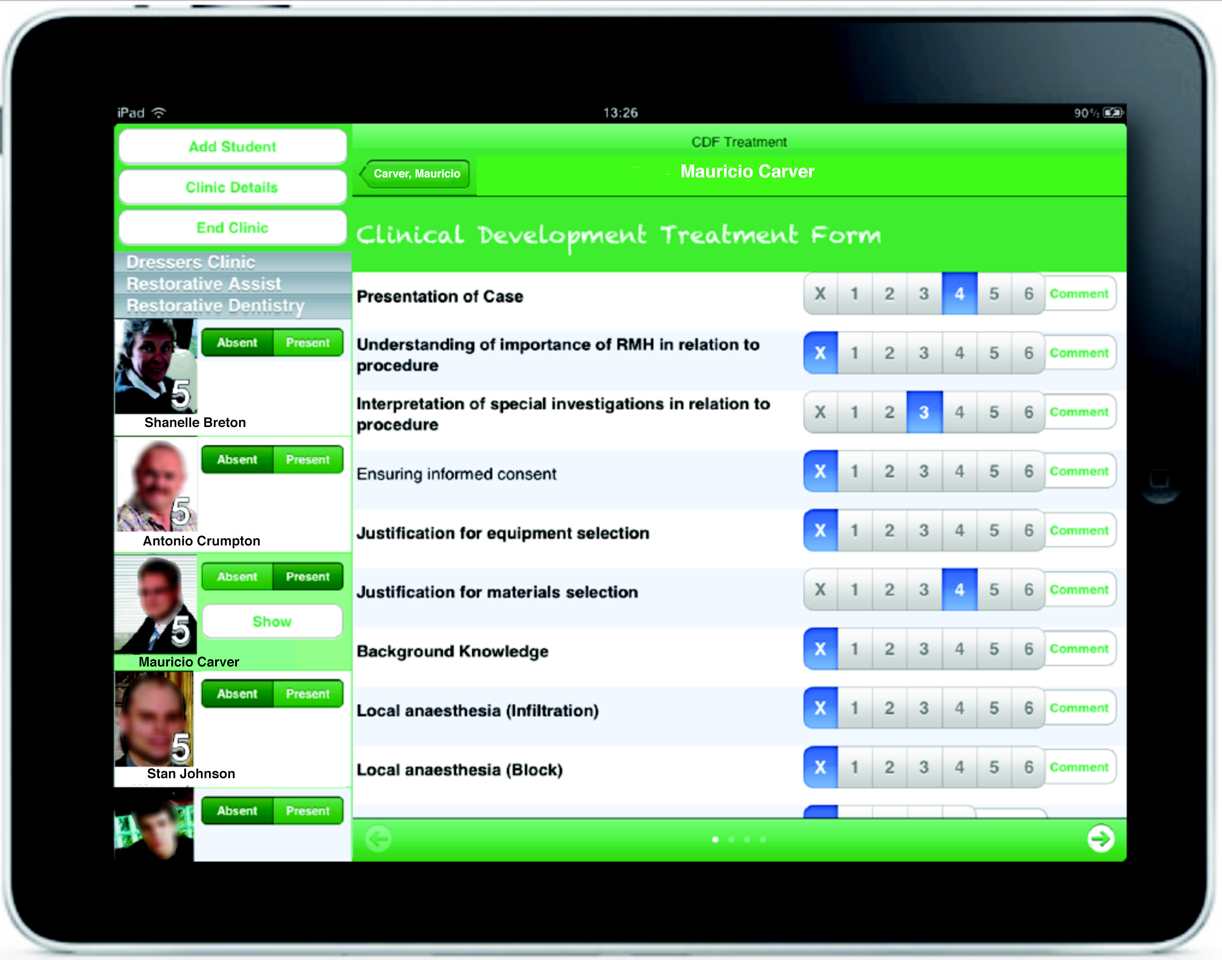} 
\par\end{centering}

\label{fig:ipad} }~~~\subfloat[\emph{Barcodes} visualise consistency.]{\begin{centering}
\includegraphics[width=6cm]{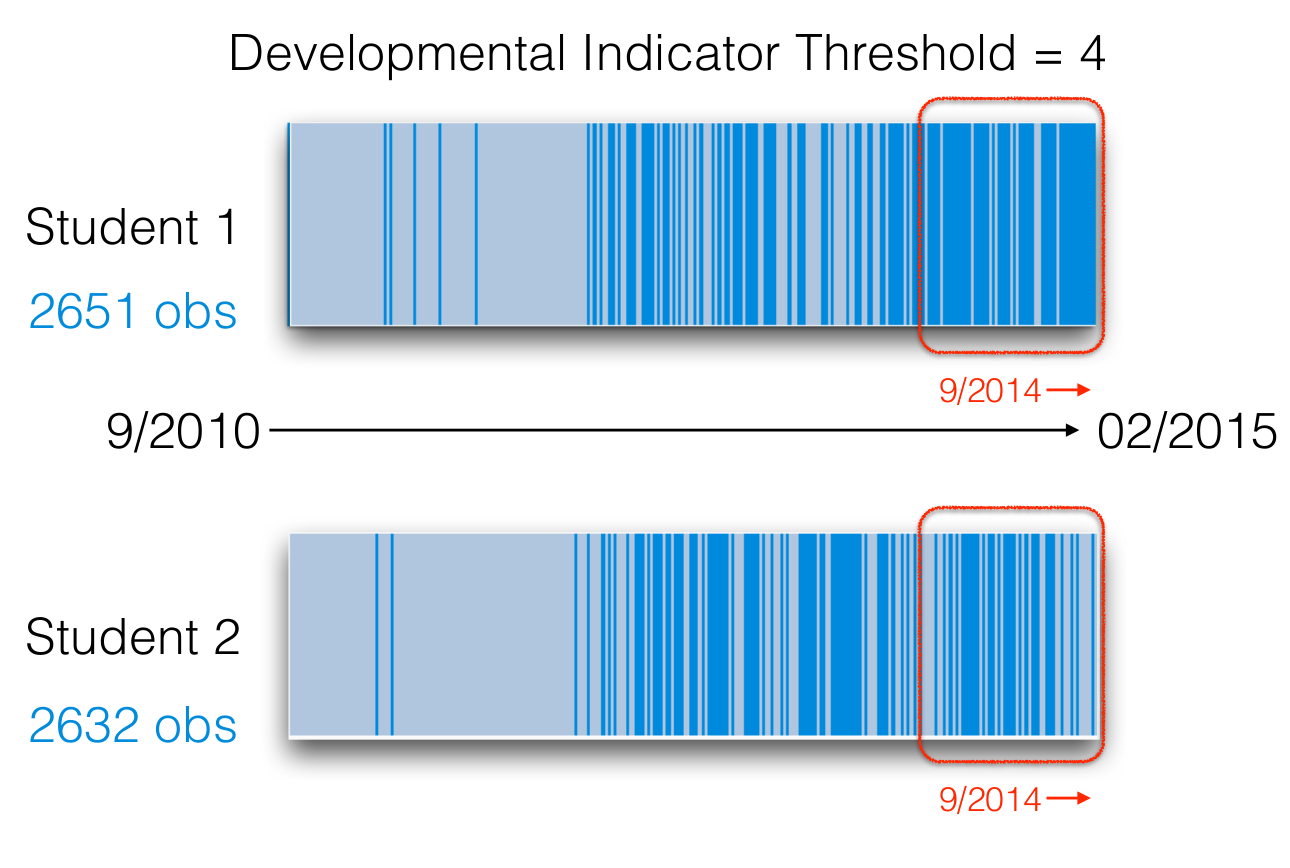}\negthinspace{}\negthinspace{}\negthinspace{}\negthinspace{}\negthinspace{}\negthinspace{}\negthinspace{}\negthinspace{}\negthinspace{}\negthinspace{}\negthinspace{}\negthinspace{}\negthinspace{}\negthinspace{}\negthinspace{}\negthinspace{}
\par\end{centering}

\label{fig:barcode} }\protect\caption{Collection and visualization of data.}
\end{figure}
The above work flow based approach was initially rolled out using
paper forms that were optically read. This was because, in 2009,
mobile technology was not up to the task, and due to its real-time
observational nature in the clinic desktop computers were considered
as to be a barrier to the process. While this worked in principle,
the amount of administrative overhead was huge. At the time 80,000
paper forms per year were being generated, which lead to around a
10\% data loss, delays in data upload, and made it impossible for
students to see any written comments that staff wished to provide
to enhance their performance. In 2010, following the release of the
iPad we could operationalise the process as intended. 

The app, shown in Fig.~\ref{fig:ipad}, is tailored to deal with
all possible work flows and is designed for easy navigation during
observations. For instance, it allows for rapid switching between
students, and the main part of the screen is designed to simultaneously
display many of the workflow items. In addition, the app is aware
of its location so the correct work flows can be automatically selected,
the right students can be assigned in order to allow simple monitoring
of their attendance, and the right staff can sign in, or make the
system aware they are covering for a colleague. In addition, the app
was designed to work-off line as Wi-Fi access is not constant in some
clinical areas. At the end of the session each student is provided
with personalised feedback on the device, and then signs out to confirm
they have seen it, at which point their record is locked. Following
the sign out of all students, the staff member signs out and the data
it uploaded to the core.

\section{Data Fusion, Visualization \& Use}

\label{sec:data-fusion-and-use}While we believe that advanced AI
techniques have the potential to radically improve the development
of performance in students, the current system already benefits from
the collected data in various ways. Here, we review these, thus underlining
how LiftUpp makes a first step towards data-supported improvement
of dentistry and, more generally, workplace-based education.

\paragraph{Quality Assurance for Curriculum and Assessment Design.}

The fact that all exam questions and work flow items are coupled to
learning outcomes, means that is it is possible to automatically verify
if the requirements of accreditation bodies are satisfied. In fact,
LiftUpp takes this one step further by providing an interface to semi-automatically
generate exams meeting those requirements. In addition, the exam question
database can also be used (e.g., by external examiners) to check the
quality of the questions. Furthermore, the collected data from taken
exams can be used to track the performance on individual questions,
thus allowing for further improvements.

\paragraph{Progress Monitoring.}

One of the greatest challenges is to take the complex data and display
it in a manner that is simple for both the learner and the staff to
understand. With respect to the presentation of the clinical performance
data the challenges are significant: How do you display 5061 data
points from hundreds of clinical sessions, covering 30 skills, in
a meaningful way? We have found that, without support, staff resort
to \textquoteleft counting\textquoteright{} the numbers of procedures
and problems. This is far too simplistic: we need to judge performance
over time. Students should not only attain a level of performance,
but also \emph{consistently }perform at that level afterwords. 

We have not fully resolved this problem, and work is ongoing, but
a major step forwards was the definition of what we have termed \textquoteleft sessional
consistency\textquoteright : the fraction of a student's sessions
that meets a desired performance threshold level. The logic to doing
this is that undertaking the job of dentistry is complex, and each
skill is interdependent. Therefore, in any area where any skill falls
below the required level, the ability to do the job is affected. To
represent this sessional consistency visually we developed the `barcode'
view. Fig.~\ref{fig:barcode} shows two examples of barcodes. They
display each observation that meets the required level of performance
(in this case, it was assessed as 4 or higher) in blue. The top barcode
conveys that the student has learned to perform consistently well
over the last period of time, while the bottom barcode corresponds
to a student that is able to reach the required level, but does not
yet do so consistently. As more data becomes available, it will also
be possible to  have better expectations of the development in the
early stages of studies, thus allowing for better early detection
of problems and advice to students.

\paragraph{Adaptive Instructional Planning.}

The ability to better monitor student progress is also useful for
creating adaptive schedules for students. The patients that require
a particular treatment at any given time are a limited resource, and
as such the allocation of students to patients is an important question
in a dental school. The insight that LiftUpp provides about the students
performance is used to decide which students will benefit most from
additional practice opportunities, while other students are put in
`holding patterns', which means that the frequency of their workplace-based
assessment is reduced or shifted towards less resource-limited treatments.

\paragraph{Feedback for Students.}

\begin{figure}
\subfloat[Feedback to students in the form of their portfolio.]{\begin{centering}
\includegraphics[width=0.6\columnwidth]{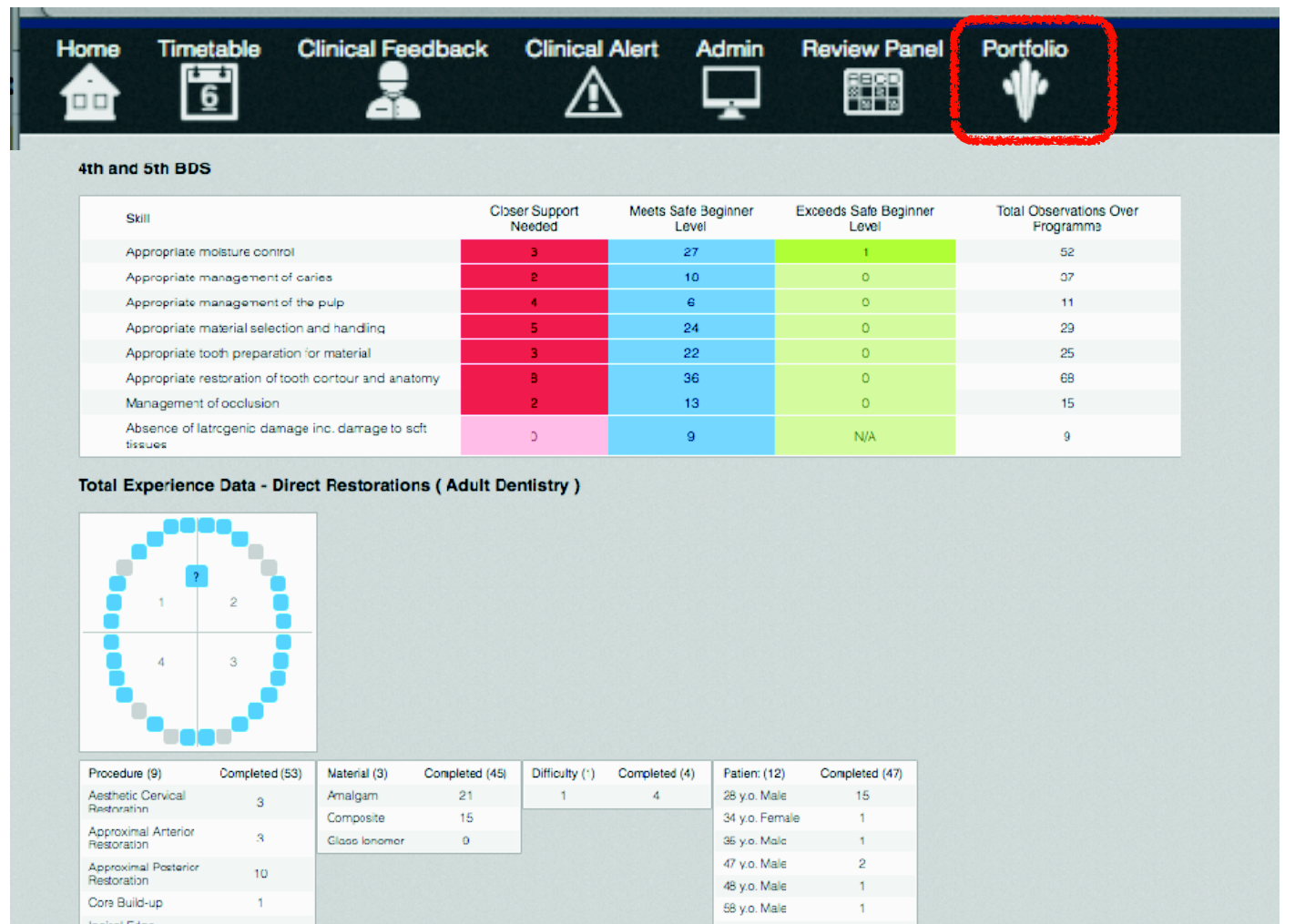} 
\par\end{centering}

\label{fig:portfolio}

}~~~~\subfloat[Slicing over staff for feedback on assessment.]{\begin{centering}
\includegraphics[width=0.3\columnwidth]{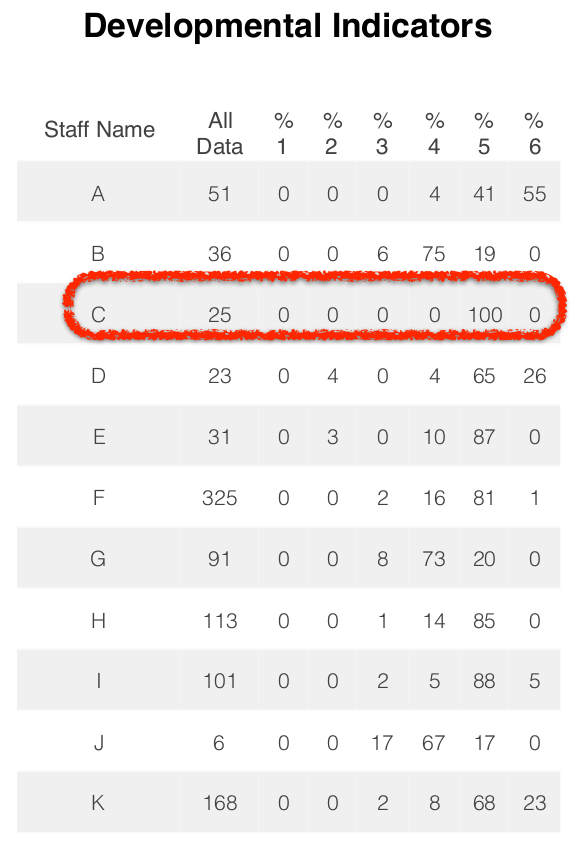} 
\par\end{centering}

\label{fig:staff-feedback}

}

\protect\caption{Some other ways in which LiftUpp facilitates providing feedback.}
\end{figure}

The bar codes are not only used for progress monitoring, but can also
be seen by students, to inform their holistic development. The students
can investigate their own performance in detail through the additional
information provided: they can drill down to the level of each individual
skill and see both the contextual performances and their experience.
Additionally, the interface presents the student with a view of their
`portfolio', shown in Fig.~\ref{fig:portfolio} which gives insight
in what procedures are sufficiently demonstrated and which need more
training. This produces a meaningful transferable employability profile,
able to inform employers over the individual learner\textquoteright s
strengths and weakness, so ongoing graduate training can also be tailored.
\begin{figure}[t]
\begin{centering}
\includegraphics[bb=0bp 100bp 932bp 623bp,clip,width=1\columnwidth]{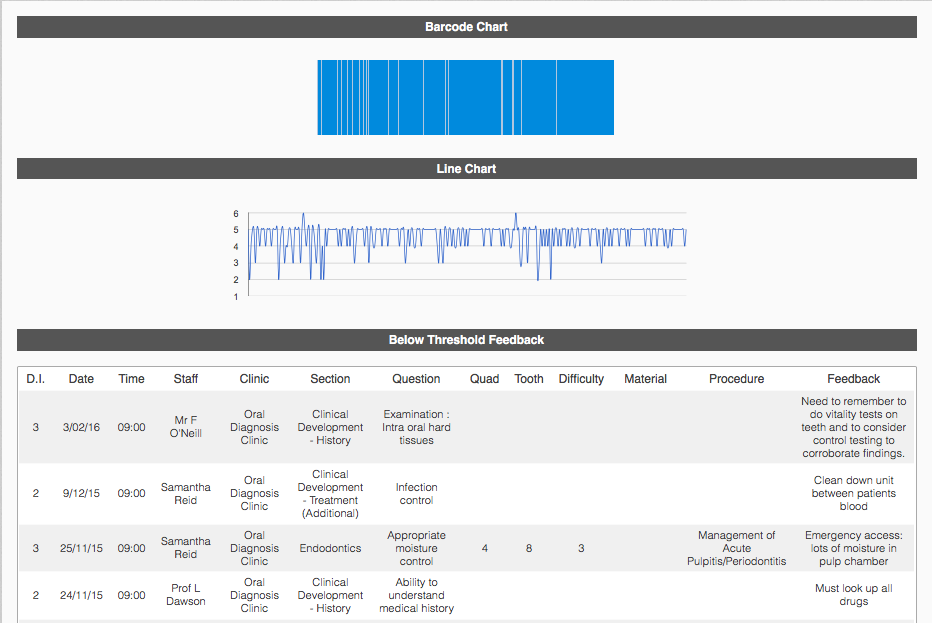}
\par\end{centering}

\protect\caption{Student feedback in LiftUpp.}

\label{fig:student-feedback}
\end{figure}

\paragraph{Feedback for Observers.}

A further function that is allowed by the LiftUpp design is the ability
to explore staff calibration, since the collected data can be sliced
by member of staff. The proper use of the developmental indicators
is essential for the credibility and trustworthiness of the data,
but also requires training. Well-trained staff use the full range
of the developmental indicator scale, but some staff may only use
2 or 3 out of 6 points on the scale. This means that their \textquoteleft 3'
may be equivalent to someone else's \textquoteleft 2\textquoteright .
The LiftUpp can give staff members insight into indicators for which
their assessment deviates from other members of staff, as illustrated
in Fig. \ref{fig:staff-feedback}. \textbf{} This information can
be used to train observers.

\paragraph{Defensible Decisions at Board of Examiners.}

A final way in which LiftUpp currently uses the collected data is
by supporting the decisions at the board of examiners. Defensibility
in clinical programmes is a big issue because we are talking about
ending someone's career before it even starts. The potential lifelong
earnings of a dentist have been estimated to be £3.5M, and this is
what the school could be sued for; a wrong decision over student progression
can be catastrophic for the institution. The defensibility of any
judgement is ultimately related to the quality of the data collected,
and how it is interpreted, and on these fronts LiftUpp has significantly
pushed the envelope.

\begin{figure}[t]
\includegraphics[bb=0bp 380bp 1640bp 1079bp,clip,width=12cm]{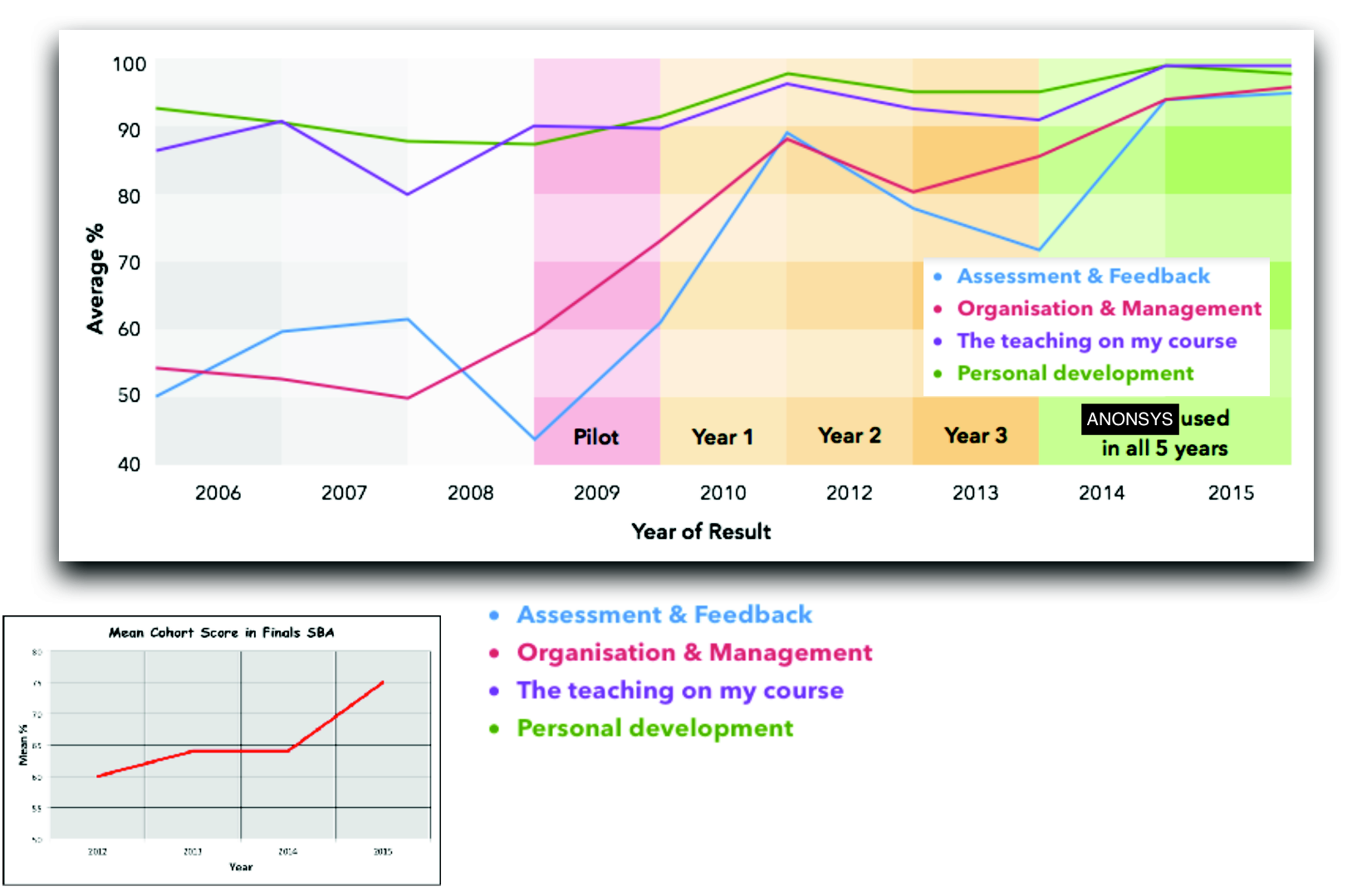}

\protect\caption{Student satisfaction ratings since the introduction of LiftUpp.}

\label{fig:NSSratings} 
\end{figure}

\section{Deployment and Effectiveness}

LiftUpp is a system that is developed in a deployed context; it has been in use
in the School of Dentistry at the University of Liverpoool since 2009, and has
made an unprecidented impact on the student experience, and their learning. This
is perhaps most clearly expressed by the student satisfaction ratings (as
measured by the national student satisfaction (NSS) survey), shown in
Fig.~\ref{fig:NSSratings}. The figure clearly shows that there are improvements
over all four categories, but that especially the satisfaction with respect to
assessment \& feedback as well as organization \& management have markedly
improved since the introduction of LiftUpp.

The breadth and quality of the collected data is high and has been
key in running a more effective administration. We estimate that LiftUpp
has saved approximately £150,000 in administration costs. Moreover,
the data has been used successfully in several legal cases where students
have challenged decisions, all the way to the Office of the Independent
Adjudicator and General Dental Council.

This success has caused other dental schools to take interest and
LiftUpp is now deployed in 70\% of UK dental schools. Moreover, in
recent years there has also been interest from other workplace-based
disciplines, leading to deployments in veterinary medicine, physiotherapy,
nursing and other healthcare courses.

\section{Conclusions \& Future Research Directions}

In this paper, we presented some of the challenges addressed by LiftUpp
for supporting the development of learners' performance in dentistry.
The primary motivation for initiating the development for LiftUpp
came from the need to comply with regulations and improve student
satisfaction. However, in doing so, LiftUpp has to a large extent
addressed the difficulties that one faces in terms of data collection
when trying to apply data-driven approaches as developed in the AIED
community to educational programs based on workplace based assessment.
In addition, LiftUpp makes some first but effective steps in dealing
with the resulting data fusion problem for a variety of uses ranging
from quality assurence, to various forms of feedback and instructional
planning. Moreover, the years in which the system has been employed
has now generated a wealth of data that may serve as the basis for
better data fusion techniques and thus form the basis of many future
directions of research. The three most promising directions we describe
in the remainder of this section.

\paragraph{Advanced Statistical Methods for Data Fusion, Interpretation and
Calibration. }

An important direction of future work will focus on investigating
the applicability of statistical and machine learning methods to better
interpret the available data. For instance, the bar codes and consistency
measures discussed in Section \ref{sec:data-fusion-and-use}\textbf{
}have been very useful in supporting the current practice, but also
have shortcomings; e.g., if you calculate the consistency of individuals,
and several students score 81\%, this hardly implies that these students
are at the same level of performance: the observations below the threshold
could relate to minor or serious issues, and their distribution over
time could be very different. Therefore, we need mechanisms to look
at the patterns, and their causes, to help inform staff of whose progress
is on schedule and whose might pose a problem.. This line of work
could lead to novel definitions of metrics such as consistency of
learner performance that are better supported by the available data
themselves. Another interesting question here is whether it is possible
to come up with better methods for calibration of the interrelated
strictness of different teachers, hardness of different tasks, and
skill of different students.

\paragraph{Adaptive Tutoring.}

In its current form, LiftUpp already supports some amount of adaptivity
in the roster of the students as discussed earlier in Section~\ref{sec:data-fusion-and-use}.
Future research, however, could focus on building on intelligent tutoring
systems research to more intelligently select next procedures to perform.
Additionally, such adaptive elements could also be supported at a
much finer level. For instance, one can conceive of an approach where
the system knows what the student has been doing in the clinic and
then intelligently issues a series of questions through adaptive testing
approaches that contextualise the knowledge, which the student is
required to answer there and then in the clinic. Over time, this would
reinforce the association between the knowledge base and its practical
application.

\paragraph{Personalized Feedback and Student Advice.}

At a coarser level, such approaches will be useful to assist in advising
students. Staff can be variable in their ability to give feedforward
advice. Good advice is predicated around sets of principles that are
evidence based. Through analysis of student performance data we think
it will be possible to give well constructed development advice personalised
to the learner. Similarly, one can imagine the use of such data analyses
for  advising over the need for a trajectory change, or the need to
experience particular contexts. This is essential for the development
of self-regulated learning. But in the extreme, since the cost to
the tax payer per dental student is £250K, it is better for both the
students and public to know early if they are not going to make it.

{\small\bibliographystyle{abbrv}
\bibliography{references}
}
\end{document}